\documentclass{article}

\usepackage{graphicx}
\usepackage{epsfig}

\def\abstract#1{\vskip 7mm 
        \begin{center}{\large Abstract}\par \smallskip
                \begin{minipage}[c]{12cm}
                        \small #1
                \end{minipage}
        \end{center}
}
\def\title#1{\begin{center}{\Large\bf #1}\end{center}}
\def\author#1{\vskip 5mm \begin{center}{#1}\end{center}}
\def\address#1{\begin{center}{\it #1}\end{center}}
\makeatletter
\@ifundefined{lesssim}{}{}
\@ifundefined{gtrsim}{}{}
\def\vereq#1#2{\lower3pt\vbox{\baselineskip1.5pt \lineskip1.5pt
\ialign{$\m@th#1\hfill##\hfil$\crcr#2\crcr\sim\crcr}}}
\makeatother

\begin{document}

\title{Radionic Non-uniform Black Strings}
\author{Takashi Tamaki\footnote{E-mail:tamaki@tap.scphys.kyoto-u.ac.jp},
Sugumi Kanno\footnote{E-mail:sugumi@tap.scphys.kyoto-u.ac.jp} and 
Jiro Soda\footnote{E-mail:jiro@tap.scphys.kyoto-u.ac.jp}  }
\address{Department of Physics, Kyoto University,
606-8501, Japan }

\abstract{
  Non-uniform black strings in the two-brane system are investigated 
using the effective action  approach. 
It is shown that the radion acts as a non-trivial hair of black strings.
The stability of solutions is demonstrated using the catastrophe theory. 
The black strings are shown to be non-uniform. }

\section{Introduction}

In the Randall-Sundrum 1 model~\cite{tama-r6Randall}, 
 the black hole can be regarded as a section of the black string as long as
 the distance between two branes is less than the radius of the black hole 
 on the brane. As the radion controls the 
length of the black string, it can trigger  the transition 
from the black string to localized black hole through the Gregory-Laflamme
 instability. The purpose of this work is to reveal the role of the radion
 in the black string system with the hope to understand this phenomena.
 We take the specific model that  the dilaton field 
coupled to the electromagnetic field on the $\oplus$-brane.
 In the case of stable black string, we can use the low energy approximation
that the curvature on the brane is 
smaller than the curvature in the bulk. Foutunately,
 the effective action is known in this case as~\cite{tama-r6Kanno}  
\begin{eqnarray}
S_{\rm\oplus}&=&\frac{1}{2 \kappa^2} \int d^4 x \sqrt{-h} 
	\left[ \Psi R (h) - \frac{3(\nabla\Psi)^2}{2(1- \Psi )}\right] 
-\int d^4 x \sqrt{-h}\left(\frac{(\nabla\phi)^2}{2}
+\frac{e^{-2a\phi}}{4}F^{2}\right)  ,
      	\nonumber 
\end{eqnarray}
where we defined $\Psi :=1-\exp (-2d/\ell )$. Here, $d$ and $\ell$ are 
the proper distance between the branes and the curvature radius in AdS$_5$, 
respectively. The point is that the bulk metric is completely 
determined by the 4-dimensional theory through 
the holographic relation as~\cite{tama-r6Kanno} 
\begin{equation}
  	g_{\mu\nu} = (1-\Psi )^{y/\ell} \left[
       		h_{\mu\nu} (x) + g^{(1)}_{\mu\nu} (h_{\mu\nu} , \Psi, 
        	T^\oplus_{\mu \nu } ,  T^\ominus_{\mu \nu } , y)  \right] \ . 
        	\label{holograms}
\end{equation}
Here, we have used the following 
coordinate system to describe the geometry of the brane model: 
\begin{equation}
ds^{2}=e^{2\eta (x^{\mu})}dy^{2}+g_{\mu\nu}(y,x^{\mu})dx^{\mu}dx^{\nu}\ . 	
\label{coordinate}
\end{equation}
We place the branes at $y=0$ ($\oplus$-brane) and $y=\ell$ ($\ominus$-brane) 
in this coordinate system. 
Using this fact, we have investigated the bulk geometry of this system and
found stable non-uniform black strings for which the radion plays
an important role~\cite{tama-r6Tamaki}. We consider the static and 
spherically symmetric metric in the Einstein frame on the $\oplus$-brane as 
\begin{eqnarray}
ds^{2}=-f(r) e^{-2\delta(r)}dt^{2}+f(r)^{-1}dr^{2}
+r^{2}d\Omega^{2}, 
\label{eqn:metric}
\end{eqnarray}
where $f(r):=1-\kappa^2 m(r)/4\pi r$. 
We assume the existence of a regular event horizon at $r=r_H$ and asymptotically 
flatness. We consider solutions with a magnetic charge $Q_m$. 
In the following, we provide views both from 
 the brane and from the bulk.

\section{View from the Brane}

\subsection{Radion as a Hair}

We obtain the first law of black hole thermodynamics, 
from which we can find that the radion field is a hair of black strings. 
We can prove that $\Psi$ is monotonically 
decreasing function of $r$. 
In Fig.~\ref{Figart3} (a), as an example, $\Psi$  for a solution 
$\bar{a}:=\sqrt{2}a/\kappa=\sqrt{3}$, $\Psi_{\infty}:=\Psi (\infty)=0.25$ 
and the horizon radius $\lambda_{H}:=\sqrt{2}r_{H}/\kappa Q_{m}=0.119$ 
is depicted. 

\subsection{Stability of the Black String}

\begin{figure}[htbp]
\psfig{file=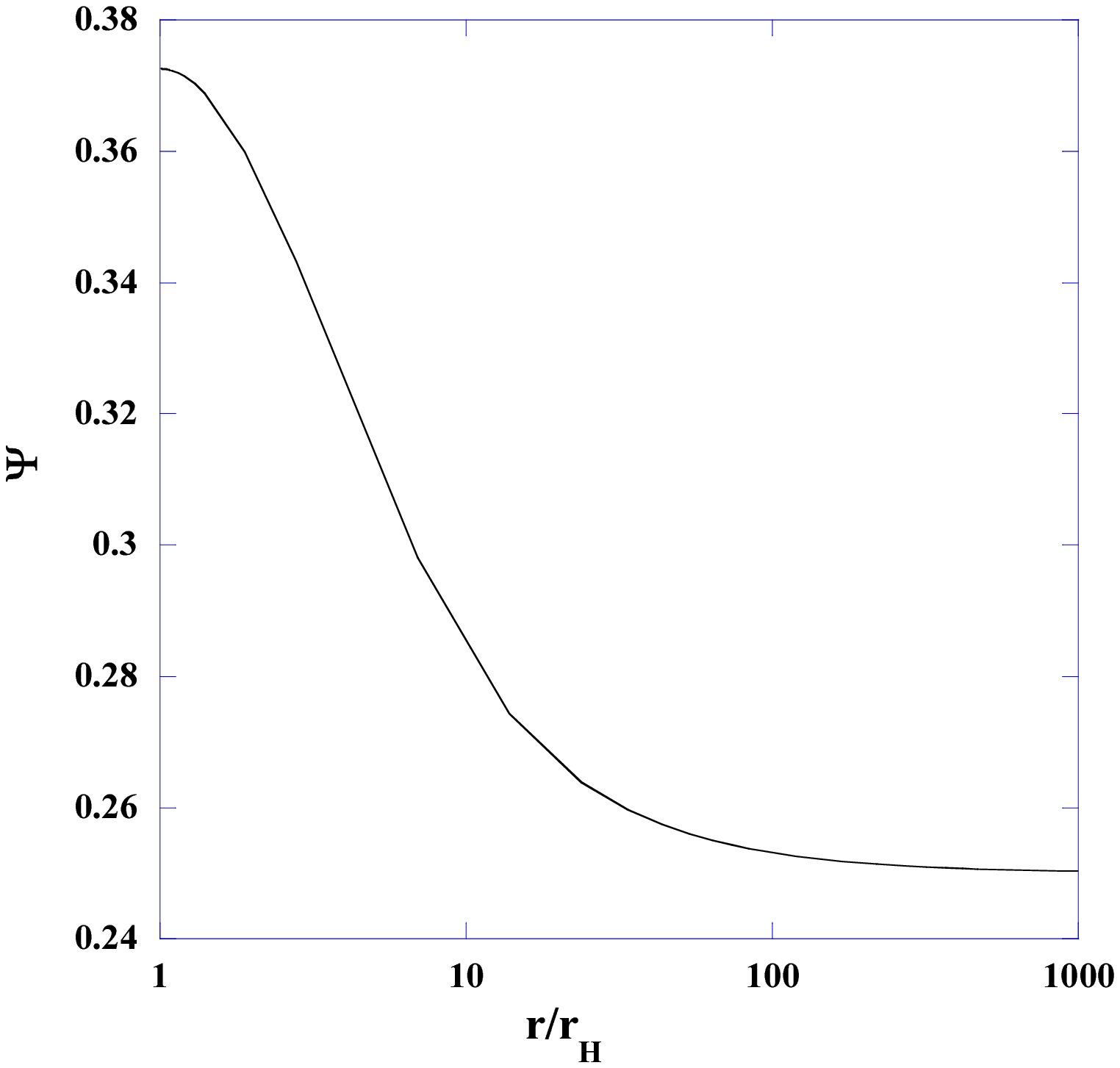,width=3.5in}
\psfig{file=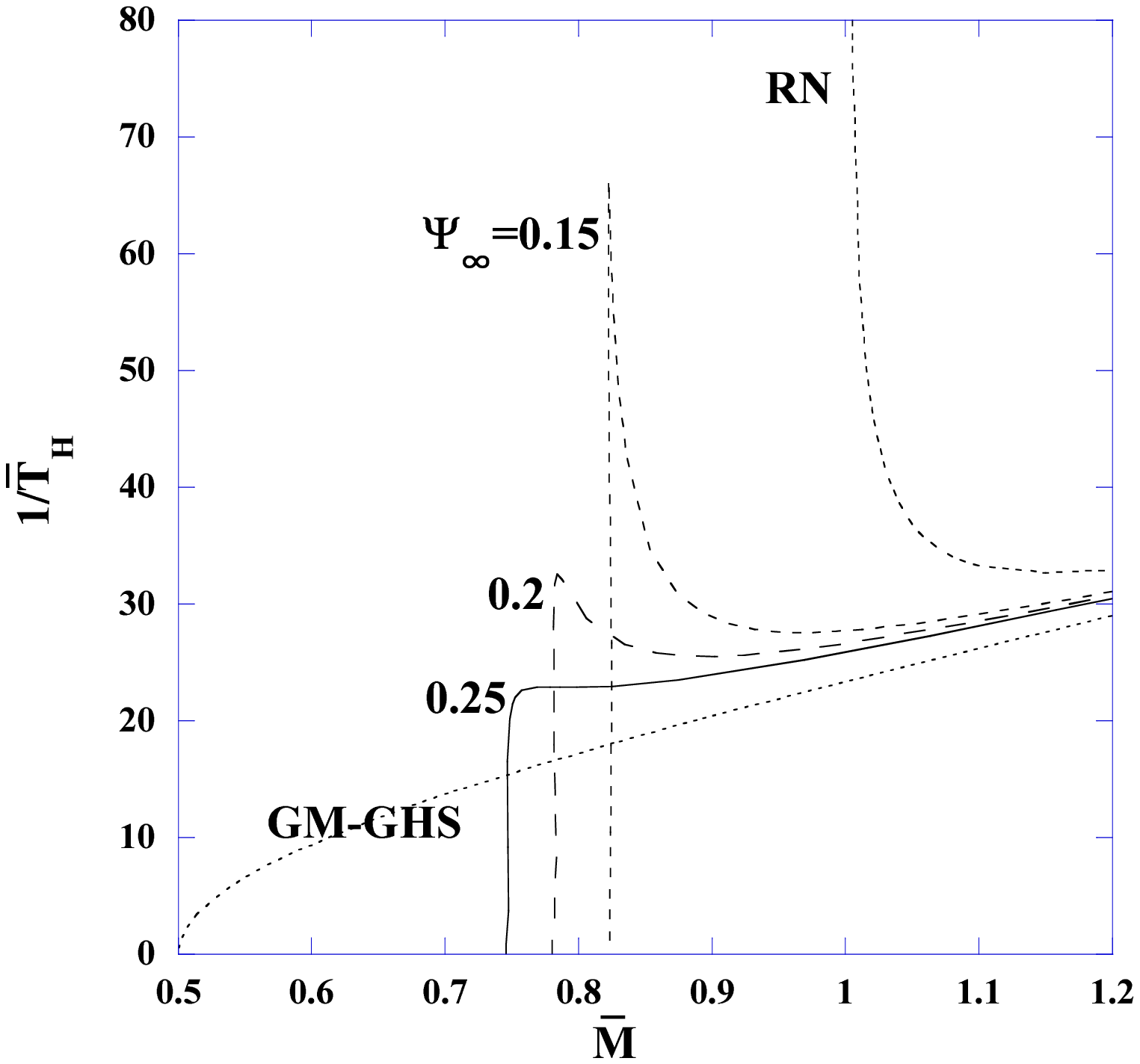,width=3.5in}
\caption{(a) The behavior of $\Psi$ (b) $\bar{M}$-$1/\bar{T}_{H}$. 
\label{Figart3} }
\end{figure}

In the single brane limit $\Psi\to 1$, the effect of the radion ceases.
 Hence, the solution approaches GM-GHS solution~\cite{tama-r6GM-GHS}. 
In the close limit $\Psi\to 0$, we have $\phi (r)=0$. 
Thus, only Reissner-Nordstr\"om (RN) solutions are possible in this limit 
 because of the no-hair theorem. 
Therefore, the non-trivial radion ($\Psi\neq 0,1$) interpolates RN and GM-GHS 
solutions as in Fig.~\ref{Figart3} (b) where the relation between the inverse temperature 
$1/\bar{T}_{H}$ and $\bar{M}$ are plotted. 

We can prove that there is no inner horizon for $\Psi\neq 0$. 
Thus, the causal structure  is the same as that of Schwarzschild black hole. 
This suggests that our solutions are stable as the GM-GHS solutions. 
We can also argue stability of our solutions using the catastrophe theory.    
According to the catastrophe theory, the stability changes at 
$d(1/T_{H})/dM=\infty$~\cite{tama-r6Katz,tama-r6catas}. 
Since we cannot find the point $d(1/T_{H})/dM=\infty$ in the graph 
for various parameters,  our solutions are stable in the catastrophic sense. 

\section{View from the bulk}

Our interest here is the $y$-dependence of the horizon which had not been 
investigated so far in the RS1 model. The hologram (\ref{holograms}) 
can be used to see the shape of the black strings.  
To see the non-uniformity of the horizon, let us investigate the change of the 
circumference radius along $y$ direction. 
The procedure to obtain a circumference radius of the horizon is summarized 
as follows: (i) seek for the radius $r=r_{+}(y)$ which satisfies 
$h_{00}+g_{00}^{(1)}=0$ 
(or equivalently $h_{11}+g_{11}^{(1)}=\infty$ as we denote below.) 
(ii) evaluate the circumference radius in the Einstein frame as 
\begin{eqnarray}
R_{H}:=\sqrt{r_{+}^{2}+\Psi (r_{+})g_{22}^{(1)}(r_{+})}\ .
\label{rh-bulk}
\end{eqnarray}
Note that we subtracted the effect of the AdS background in 
the above expression (See, Eq.~(\ref{holograms})). 

First, let us find $r_{+}(y)$. 
Writing $g_{00}^{(1)}=:h_{00}f_{0}(r,y)$ and $g_{11}^{(1)}=:h_{11}f_{1}(r,y)$, 
we can verify that $f_{0}=f_{1}$ and 
they have finite values at $r=r_{H}$. 
Hence, the coordinate value of the horizon does not change 
$r_{+}(y)=r_{H}$ even at this order.  Thus, we can evaluate the 
circumference radius (\ref{rh-bulk}) using the expression   
\begin{eqnarray}
g_{22}^{(1)}(r_{H})=-\frac{\ell^{2}w}{2}
\left[R_{22}-\frac{1}{6}h_{22}R
+(w+2)\chi_{22}\right]\ .
\label{correction22}
\end{eqnarray}
Here, we introduced the variable $w:=(1-\Psi)^{-y/\ell}-1$ which increases 
toward the $\ominus$-brane. 

\vspace{-1cm}

\begin{figure}[htbp]
\psfig{file=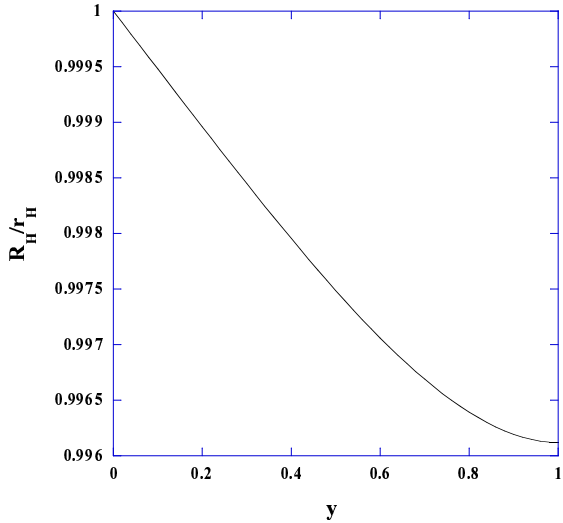,width=3.5in}
\caption{Deformation of the horizon for $\Psi_{\infty}=0.25$, 
$\lambda_{H}=1.11\times 10^{-2}$, $\bar{a}=\sqrt{3}$ and $\ell/r_{H}=0.5$. 
\label{bulk}}
\end{figure}

As an example, we show the ratio of the horizon $R_{H}/r_{H}$ as a function of 
$y$ in Fig.~\ref{bulk}. We find that the circumference radius of the horizon 
monotonically shrinks 
toward the $\ominus$-brane even if we subtract the background effect. 
Although, we showed only one example, we can easily confirm that this is 
general by differentiating (\ref{correction22}) with respect to 
$w$ since the fields satisfy the conditions 
\begin{eqnarray}
T_{22}=\frac{\Psi Q_{m}^{2}}{2r^{2}}e^{-2a\phi}>0\ ,\ \   
T^{\mu}_{\ \mu}=0\ ,
\end{eqnarray}
at $r=r_{H}$. 
In fact, by rewriting Eq. (\ref{correction22}) as 
\begin{eqnarray}
K:=-\frac{2}{\ell^{2}}g_{22}^{(1)}(r_{H})
=w\left[\frac{h_{22}}{3}R+\frac{\kappa^{2}}{\Psi}T_{22}
\left\{1-\left(w+2\right)\frac{1-\Psi}{2}\right\}\right]\ ,
\label{correction22-general}
\end{eqnarray}
we obtain 
\begin{eqnarray}
\frac{dK}{dw}&=&\frac{h_{22}}{3}R+\frac{\kappa^{2}}{\Psi}T_{22} 
\left\{1-\left(w+1\right)(1-\Psi)\right\},  \\ 
\frac{d^{2}K}{dw^{2}}&=&-\frac{\kappa^{2}}{\Psi}(1-\Psi)T_{22}<0. 
\label{kmin}
\end{eqnarray}
 From Eq.(\ref{kmin}), we see
 $ dK /dw$ takes minimum value at the $\ominus$-brane. 
On the $\ominus$-brane, we have 
\begin{eqnarray}
\frac{dK}{dw}=\frac{1}{3}h_{22}R=-\frac{1}{3}h_{22}T^{\mu}_{\ \mu}=0\ .
\label{kmin2}
\end{eqnarray}
Therefore, $dK/dw$ is always positive in the bulk.
This means $ dg_{22}^{(1)} / dw <0$. As the horizon has non-trivial $y$-dependence, 
the black strings are non-uniform. 

Finally, we comment on the zeroth law of the non-uniform black string. 
Using the metric (\ref{eqn:metric}), the Hawking temperature is 
\begin{eqnarray}
T_{H}=\frac{2\pi-\kappa^{2}m_{H}'}{8\pi^{2} r_{H}}e^{-\delta_{H}}
\sqrt{\frac{1+f_{0}(r_{H},y)}{1+f_{1}(r_{H},y)}}\ ,
\label{TH}
\end{eqnarray}
Since $f_{0}=f_{1}$ at $r=r_{H}$, $T_{H}$ does not depend on $y$. 
Therefore, the zeroth law of black hole thermodynamics holds.  

\section{Conclusion }
Non-uniform black strings in the two-brane system are investigated 
using the effective action  approach.  We considered the dilaton field 
coupled to the electromagnetic field on the $\oplus$-brane. 
It is shown that the radion
acts as a non-trivial hair of the black strings. From the brane point of view,
the black string appears as the deformed GM-GHS black hole which becomes 
GM-GHS black hole in the single brane limit and reduces to the RN black hole 
in the close limit of two-branes.  In view of the catastrophe 
theory, our solutions are stable. 
From the bulk point of view, the black strings are proved to be non-uniform. 
Nevertheless, the zeroth law of black hole thermodynamics holds. 

We established the picture that the event horizon 
shrinks toward the $\ominus$-brane (even if we subtract the effect of 
the AdS background) 
and the distance between branes 
decreases toward the asymptotically flat region. 

However,  we cannot apply our present analysis if the distance between branes 
exceeds the horizon radius. This is because the Kaluza-Klein effect becomes 
significant. It is the point that Gregory-Laflamme instability commences. 
The transition to the localized black hole may occurs. 
The AdS/Conformal field theory correspondence argument suggests 
the classical evaporation of the resultant black hole. 
Moreover, there is also a possibility that the shape of the horizon 
becomes complicated. 
To get a hint, we need to proceed to the next order calculations 
corresponding to Kaluza-Klein  corrections. 
We want to investigate it in the future.

\end{document}